\documentclass[12pt,preprint]{aastex}
\shorttitle{Explosive Shaping of PPN}
\begin{document}

\title{PPN as Explosions: Bullets vs Jets
       and Nebular Shaping}
\author{Timothy J. Dennis\altaffilmark{1}, Andrew J. Cunningham\altaffilmark{1}, Adam Frank\altaffilmark{1}}
\email{tdennis@pas.rochester.edu}
\author{Bruce Balick\altaffilmark{2}, Eric G. Blackman\altaffilmark{1}}
\and
\author{Sorin Mitran\altaffilmark{3}}
\altaffiltext{1}{Department of Physics \& Astronomy, University of Rochester, Rochester, NY 14627}
\altaffiltext{2}{ Department of Astronomy, University of Washington, Seattle, WA 98195}
\altaffiltext{3}{ Department of Mathematics, University of North Carolina, Chapel Hill, NC 27599}
\begin{abstract}

Many proto-planetary nebulae (PPN) appear as narrow collimated structures sometimes showing 
multiple, roughly aligned lobes. 
In addition, many PPN flows have been shown to have short acceleration times. 
In this paper we explore whether jet or ``bullet'' (a massive clump) models fit the observations of 
individual collimated lobes adequately by comparing simulations of both radiatively cooled (stable) jets and bullets.
We find that the clump model is 
somewhat favored over jets because (1) it leads to greater collimation of outflows
(2) it accounts better and more naturally for ring-like structures observed in the PPN CRL 618, and 
(3) it is more successful in reproducing the Hubble-flow character of observed kinematics in some PPN. 
In addition, bullets naturally account for observed multipolar flows, since the likely MHD launch 
mechanisms required to drive outflows make multiple non-aligned jets unlikely.  
Thus we argue that PPN outflows may be driven by explosive MHD launch mechanisms such as those 
discussed in the context of supernovae (SNe) and gamma-ray bursts(GRB).

\end{abstract}
\section{Introduction}
\label{sec:intro}

In the past decade, images of very young planetary nebulae (PNs) and 
proto-planetary nebulae (PPNs) have revealed an unexpected diversity of morphological 
classes.  Many of these objects appear to exhibit a level of complexity that 
cannot be accounted for in terms of the Generalized Interacting Stellar Winds model 
(GISW; Balick \& Frank 2002 and references therein). Of particular interest are 
objects exhibiting point-symmetric, multi-polar, and ``butterfly'' morphologies, 
as well as bipolar and multi-polar objects exhibiting highly collimated 
``jet-like'' outflows.

The appearance of these collimated and sometimes multi-polar outflows in so many 
PPNs has led to the suggestion that high-speed jets operate during the late 
asymptotic giant branch (AGB) and/or post-AGB evolutionary phases of the central 
star (Sahai \& Trauger 1998). While the GISW model can account for narrow jets 
(Icke et al. 1992; Mellema \& Frank 1997; Borkowski, Blondin, \& Harrington 1997), 
it assumes the winds are radiatively driven.
Radiative acceleration cannot however account for these flows since a number
of observational studies 
demonstrate a momentum excess such that a factor $\sim 10^3$ exists 
between outflow momentum observed and what can be attributed to stellar 
radiation pressure (Bujarrabal et al. 2001). Moreover, it is difficult to 
attribute the degree of observed collimation to a large-scale dust 
torus as is usually required in the GISW model.  In addition, the problem of accounting 
for the precession necessary for the production of point-symmetric flows remains to be solved
[for an accretion disk based model see Icke (2003)]. For these reasons, the 
suggestion has been made that PPN jets and collimated outflows are magnetically 
driven (Blackman et al. 2001a, 2001b; Frank \& Blackman 2004; 
Matt, Frank, \& Blackman 2006, Frank 2006).  Magnetically driven models couple rotation
to a magnetic field.  Jets therefore are bound to flow along the rotational axis of
the central object and it is difficult to see how multiple jets of similar size can
be driven by such a mechanism. We discuss these models and this issue in more detail at
the end of the paper.
\par
Observationally, clumps and collimated flows occur in many stellar outflows though not 
always together. The outflows in Wolf-Rayet (WR) nebulae are clumpy, but jets are not observed. 
In young stellar objects (YSOs), jets and collimated bipolar outflows are quite common,
and while they can often be clumpy,
the jet beams---distinct from the bow shocks which they drive---are often apparent,
stretching all the way back to the stellar source. In mature PN, clumps are often seen
[as in NGC 2392 (Eskimo), 6853 (Dumbbell) and NGC 7293 (Helix)].  Fully articulated 
jets are however very rare. We note that ionization shadows and ``mass loaded'' flows behind 
clumps can give the appearance of jets. In some mature PN such as the Cats Eye nebula, 
structures appear (some of which fall under the term FLIERS) which may be the remnants of 
poleward-directed flows.  In HST images of many PPNs, the outlines of reflected light are 
often bipolar, but within these boundaries the illuminated gas seems irregular.  Thin jets 
(as opposed to thin finger-shaped lobes) are rarely seen directly except (perhaps)
in 0H231.8+04.2 (Calabash) . However, pairs or sets of knots lying along or near the 
apparent symmetry axes are not unusual (M1-92, IRAS 20028 +3910, IRAS 16594-3656, Hen 3-1475).  
Thus the creation of continious jets as in the case of YSOs does not seem to be the norm in PN and PPN. 
\par
Clumps or ``bullets'' driven into the surrounding media have been found to be an effective explanation 
for some stellar outflow structures. In Poludnenko, Frank, \& Mitran (2004) the authors modeled 
the strings of $\eta$ Car as bullets of high speed material ejected by the star.  
The simulations showed that long, thin 
morphologies similar to jets were readily obtained along with multiple rings associated with 
vortex shedding and the break-up of the clump.  The authors suggested that such ``impulsive'' 
models may be useful in PPNs as well.  Such a scenario is very different from the jet-driven 
explanation for PPN/PN.  In this paper we seek to explore the usefulness of the clump picture.
\par
Soker (2000) has analytically explored the role of jets in PNs.  In the excellent study of 
Lee \& Sahai (2003), simulations of jets as the drivers of PPNs were presented, including 
detailed comparisons with observations. Our goals in this paper are more modest. In what
follows we take a first step in the exploration of the clumps vs. jets issue by examining 
2-dimensional and 3-dimensional pairs of simulations, with each pair consisting of either a steady 
jet impinging upon a circumstellar gaseous medium, or a clump
of gas which is fired 
ballistically through the same medium along a trajectory corresponding to the direction 
of flow in the steady jet
\footnote{In a related study Raga et al.(2007) have also recently presented a model of 
the ``3D structure of a radiative, cosmic bullet flow.''}. 
Both the jet and the clump are assumed to be magnetically 
launched though no attempt to model the launch mechanism is made here, and the simulations 
are purely hydrodynamic.  For the present, we are merely interested in examining how the 
clump and jet differ in their effect on the surrounding circumstellar medium.  As we will 
show, the jet and clump models show differences which require further study, but the clumps 
provide at least as good, or better, an account of key observational characteristics.  
Given the fact that in some cases multiple outflows are seen in a single object 
(such as CRL 618), the clump model may be more plausible since what are often interpreted 
as multiple ``jets'' could instead arise naturally from the fragmentation of an explosively 
driven polar directed shell.  We note here that none of the widely accepted magnetically 
launched outflow models would create continuous multiple jets of similar or equal age 
driven in slightly different directions. 

We note also that new models of binary stars in the context of PN's
(Nordhaus \& Blackman 2006, Nordhaus, Blackman \& Frank 2007) show the
extent to which envelope ejection can be shaped by gravitational interactions.  
In Nordhaus \& Blackman 2006 Common Envelope scenarios which lead to aspherical
mass loss (including disk creation and possible MHD launching) were
articulated.  In Nordhaus, Blackman \& Frank 2007 Common Envelope 
models were explored as the source of differential rotation in the 
primary which could drive strong dynamo supported magnetic
fields.  These models showed that while single stars may,
in some cases, be able to support a strong field over AGB timescales,
binary interactions were highly effective at creating the fields needed 
to power PPN outflows at the evolutionary moment when they will be required.  
As we will see, such models provide strong theoretical
support for the scenario we argue for in this paper.

In section \ref{sec:CompInit} we provide information concerning the numerical 
methods used,  details of the jet and clump models, and a discussion of
the initial conditions. In section \ref{sec:Results} we discuss the results of 
our simulations and in section \ref{sec:Discussion} we summarize our 
conclusions.

\section{Computational Methods and 
         Initial Conditions}
\label{sec:CompInit}
We have carried out two pairs of hydrodynamic simulations (one ``medium-resolution'' 
3D pair and one ``high-resolution'' 2.5D pair). Each pair consists of a jet and
clump respectively with each parameterized to be as similar to one another as 
possible.  Specific parameter values for the jet, clump, and ambient medium for
each simulation are given in table~\ref{tab:t1}.
The simulations are performed using the AstroBEAR code which is an extension of the 
BEARCLAW adaptive mesh refinement (AMR) package for solving conservation laws.
[For a detailed description of the AstroBEAR package see section 3 
of Cunningham Frank \& Blackman (2006).] The domain is a rectangular box with 
a square cross-section and with the $x$-axis chosen to intersect the center of the 
left square face of the domain.  The origin of coordinates is placed at this point
of intersection.  The clump and jet are launched along the $x$-axis and placed 
so that their centers coincide with it.  The jet was modeled in 3D with a circular
cross-section of radius $r_{\rm j}$ (in 2.5D the jet cross-section reduces
to a line-segment of length $2r_{\rm j}$ and a thickness of one computational cell).  
The jet was launched into the domain from a set of fixed cells along the domain
boundary. To prevent the expansion of the jet inflow boundary with time, a ring of 
zero velocity and with outer radius $1.125r_0$ was maintained around the jet-launching
region.  The velocity profile of the jet was smoothed about a nominal value $v_{{\rm j},0}$
according to 
\begin{equation}
\label{eq:jetvprof}
v_{\rm j} = v_{{\rm j},0}
\left[
1-\left(1-s\right)
\left(\frac{r}{r_0}\right)^2
\right],
\end{equation}
where $s$ is a shearing parameter taking values between $0$ and $1$ and for the
jet simulation presented here is set equal to $s=0.9$.  The clump was modeled in
3D as a spherical over-density of radius $r_{\rm c} = r_{\rm j} $.  (The sphere 
reduces to a circle in 2.5D.) Its initial position in the domain is chosen so
that its center is located at the point 
\begin{equation}
\label{eq:clumpcent}
{\mathbf r}_{\rm {c},0}         (x,y,z) = (2r_{\rm c},0,0),
\end{equation}
and the density of the clump as a function of location within the clump is
\begin{equation}
\label{eq:clumpdens}
n_c(r) = n_a(r) +  n_0
\left[
1 - 
\left(
\frac{  \left|{\mathbf r}-
              {\mathbf r}_{c,0}
        \right|}
     {r_0      }
\right)^2
\right],
\end{equation}
where
\begin{equation}
\label{eq:ambdens}
n_{\rm a}(r) = {\rm min}
\left(
n_0,\frac{n_0r_0^2}{r^2}
\right),
\end{equation}
is the ambient number density profile in regions of the domain unoccupied by jet
or clump gas, and where $r^2=x^2+y^2+z^2$, $n_0$ is the nominal ambient number
density, and $r_0$ is a characteristic length taken to be equal to the jet or
clump radius.  The 3D (2.5D) simulations are carried out on a base grid with a
resolution of 6 (12) cells per jet/clump radius and with two levels of AMR
refinement providing an effective resolution of 24 (48) cells per jet/clump radius.
In all cases, radiative cooling is modeled using the atomic line cooling function
of Delgarno \& McCray (1972), and we do not attempt to follow the detailed 
ionization dynamics or chemistry of the cooling gas. Given that both models give
rise to similarly expanding shells of shock-heated gas we do not expect
this simplification to materially affect our conclusions.
\clearpage
\begin{table*}[h]
 \scriptsize
\begin{tabular}{llll}
\hline
\hline
Model & Parameter                              & Value (2.5D)            & Value 
(3D)              \\
\hline                                                                                             
\\
Jet   & Radius, $r_j$                 \dotfill & $500 {\rm\ AU}$         & $500  
{\rm\ AU}$        \\
      & Computational cells per $r_j$ \dotfill & $48$                    & $24$                    
\\
      & number density, $n_j$         \dotfill & $500 {\rm\ cm}^{-3}$    & $500 
{\rm\ cm}^{-3}$    \\
      & peak velocity, $v_{j,0}$      \dotfill & $100 {\rm\ km\ s^{-1}}$ & $100 
{\rm\ km\ s^{-1}}$ \\
      & Temperature, $T_j$            \dotfill & $200 {\rm\ K}$          & $200 
{\rm\ K}$          \\
      & Nominal ambient density, $n_a$\dotfill & $500 {\rm\ cm}^{-3}$    & $500 
{\rm\ cm}^{-3}$    \\
      & Ambient temperature, $T_a$    \dotfill & $200 {\rm\ K}$          & $200 
{\rm\ K}$          \\
      & Shear parameter $s$           \dotfill & $0.9$                   & $0.9$                   
\\
\vspace{-6pt} \\
\hline         \\
\vspace{-12pt} \\
Clump & Radius, $r_c$                 \dotfill & $500 {\rm\ AU}$         & $500 
{\rm\ AU}$         \\
      & Computational cells per $r_c$ \dotfill & $48$                    & $24$                    
\\
      & nominal number density, $n_o$ \dotfill & $500 {\rm\ cm}^{-3}$    & $500 
{\rm\ cm}^{-3}$    \\
      & velocity, $v_c$               \dotfill & $100 {\rm\ km\ s^{-1}}$ & $100 
{\rm km\ s^{-1}}$  \\
      & Temperature, $T_c$            \dotfill & $200 {\rm\ K}$          & $200 
{\rm\ K}$          \\
      & Nominal ambient density, $n_a$\dotfill & $500 {\rm\ cm}^{-3}$    & $500 
{\rm\ cm}^{-3}$    \\
      & Ambient temperature, $T_a$    \dotfill & $200 {\rm\ K}$          & $200 
{\rm\ K}$          \\
\vspace{-0.35cm}
\\
\hline
\hline
\end{tabular}
\caption{\normalsize Simulation 
                     Parameters
\label{tab:t1} }
\normalsize
\end{table*}
\clearpage
\section{Results: Morphology}
\label{sec:Results}
\begin{figure*}[ht]
\vskip 18pt
\includegraphics[angle=0,
		 width=6.0in,
                 keepaspectratio=true,
                 trim= 0 0 0 0
                 clip=true]
        {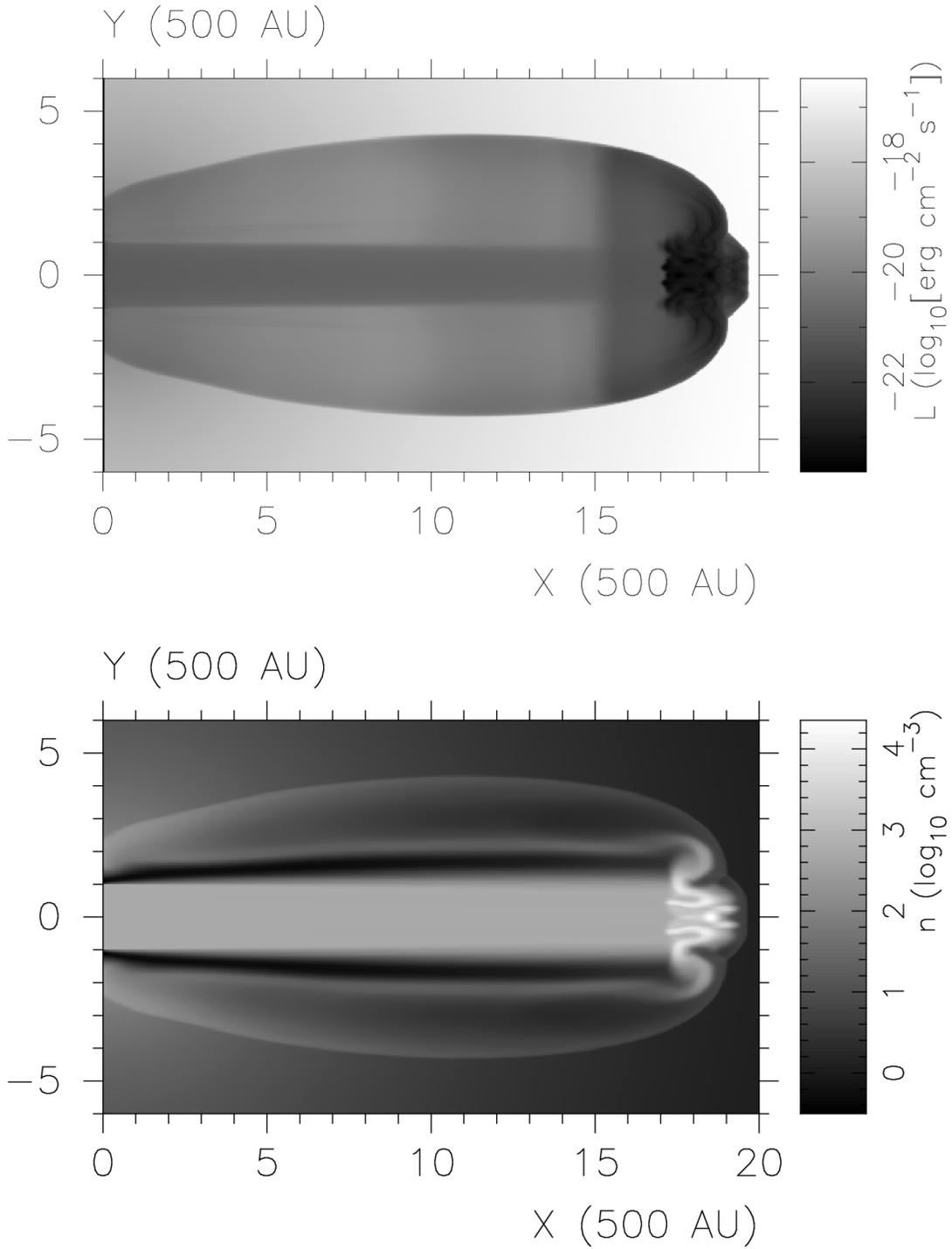}
\caption{3D Jet at time $t\simeq 636 {\ \rm yr}$,
         top: integrated emission
         assuming atomic line
         cooling; bottom, base-ten
         logarithm of density.
\label{fig:f1}}
\end{figure*}

\begin{figure*}[ht]
 \vskip 18pt
 \includegraphics[angle=0,
		 width=6.0in,
                  keepaspectratio=true,
                  trim= 0 0 0 0
                  clip=true]
        {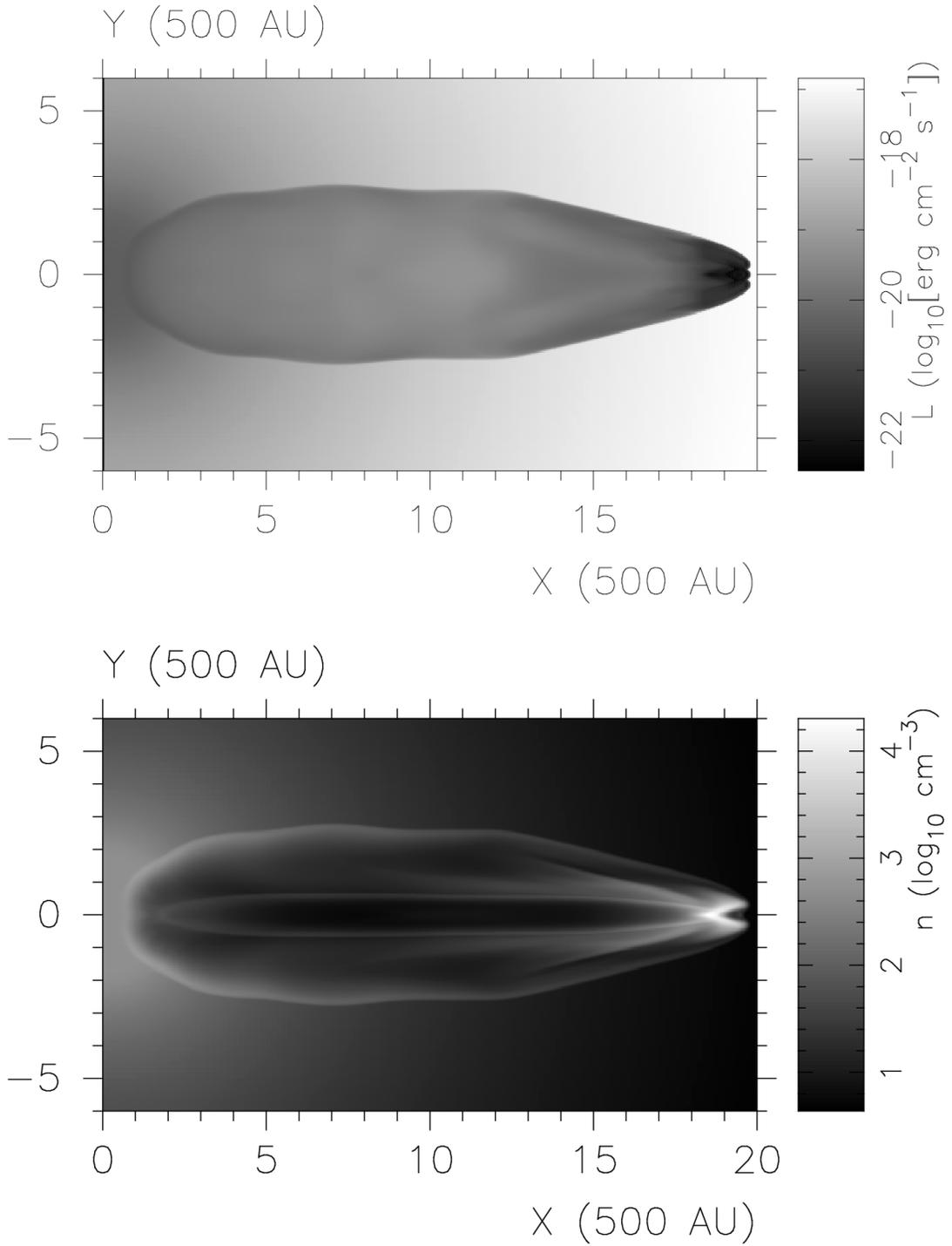}
 \caption{3D clump at time 
          $t\simeq 498 {\ \rm yr}$, top: 
          integrated emission assuming
          atomic line cooling;
          bottom, base-ten logarithm
          of density.
 \label{fig:f2}}
 \end{figure*}

\begin{figure*}[ht]
 \vskip 18pt
 \includegraphics[angle=0,
 		 width=6.0in,
                  keepaspectratio=true,
                  trim= 0 150 0 0
 		 clip=true]
         {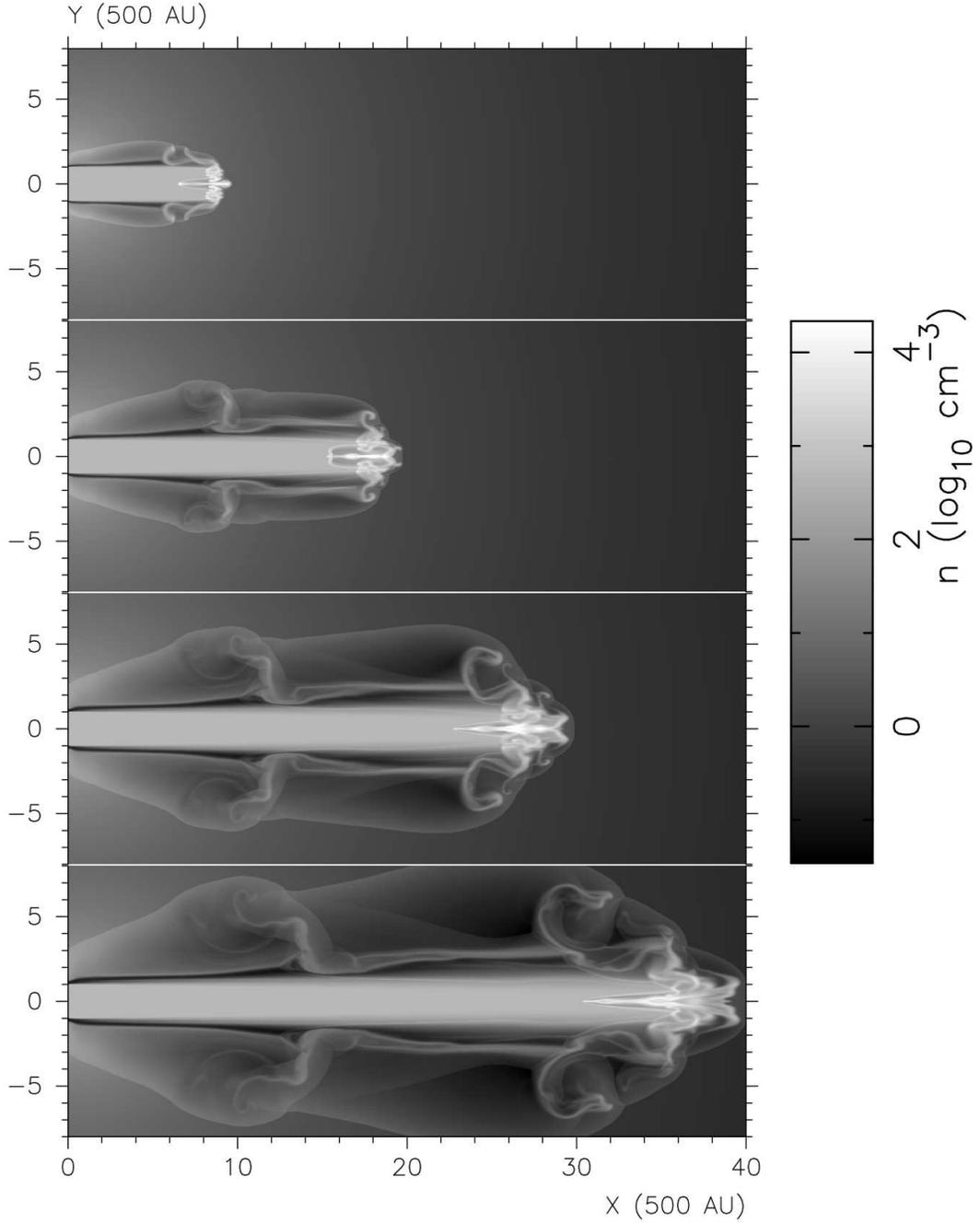}
 \caption{Density of 2.5D Jet at times
          $t\simeq 336, 628, 896,
          {\rm and\ } 1165 {\rm \ yr}$. 
          The data is shown here
          reflected about the 
          axis of symmetry.
 \label{fig:f3}}
 \end{figure*}

\begin{figure*}[ht]
 \vskip 18pt
 \includegraphics[angle=0,
 		 width=6.0in,
                  keepaspectratio=true,
                  trim= 0 150 0 0
 		 clip=true]
         {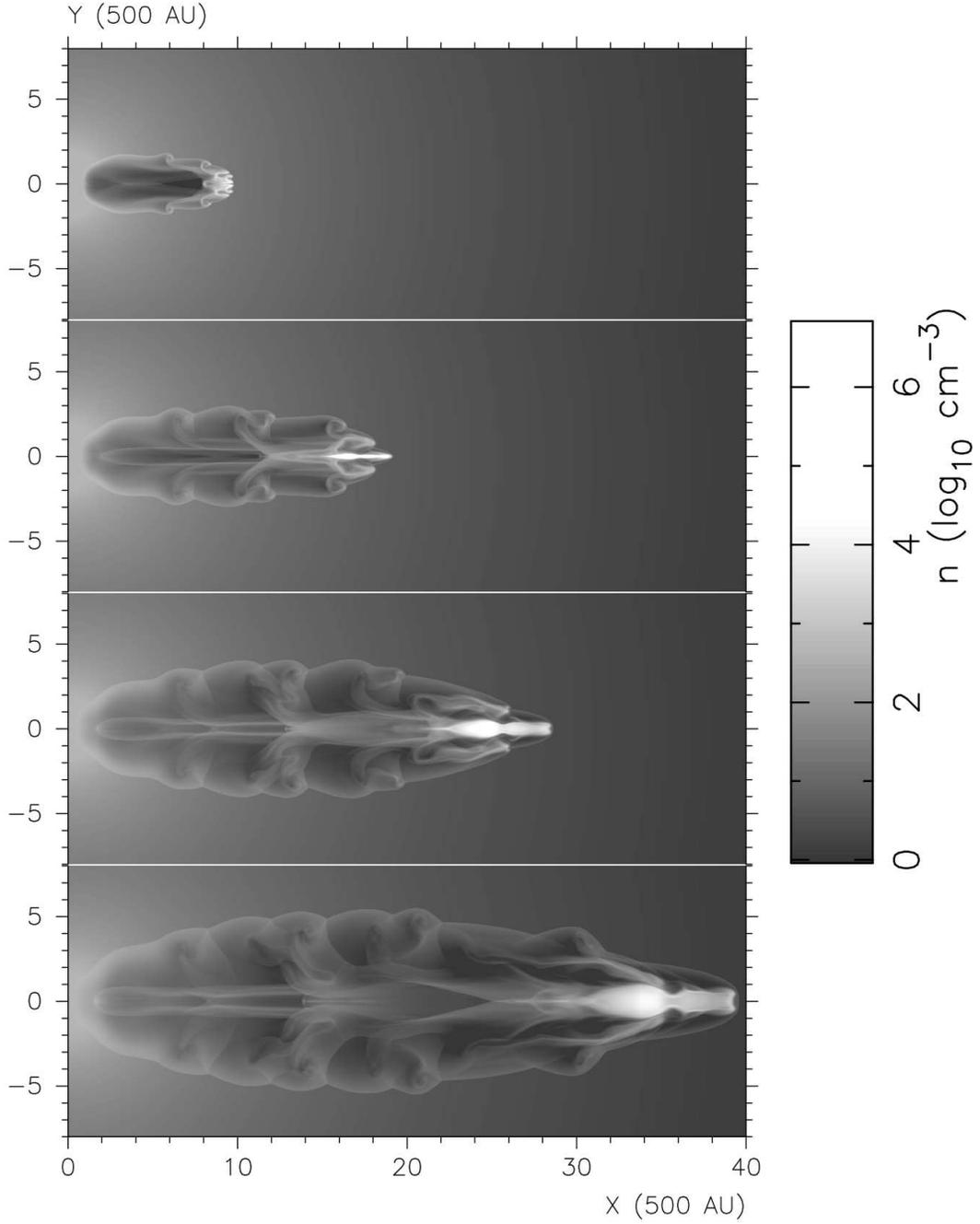}
 \caption{Density crosscut of 2.5D 
          Clump at times $t\simeq 
          208, 477, 753,{\rm and\ } 
          1082 {\rm \ yr}$. The 
          data is shown here reflected
          about the axis of symmetry.
 \label{fig:f4}}
 \end{figure*}

\begin{figure*}[ht]
 \vskip 18pt
  \includegraphics[angle=0,
 		 width=6.0in,
                  keepaspectratio=true,
                  trim= 0 300 0 0
                  clip=true]
    {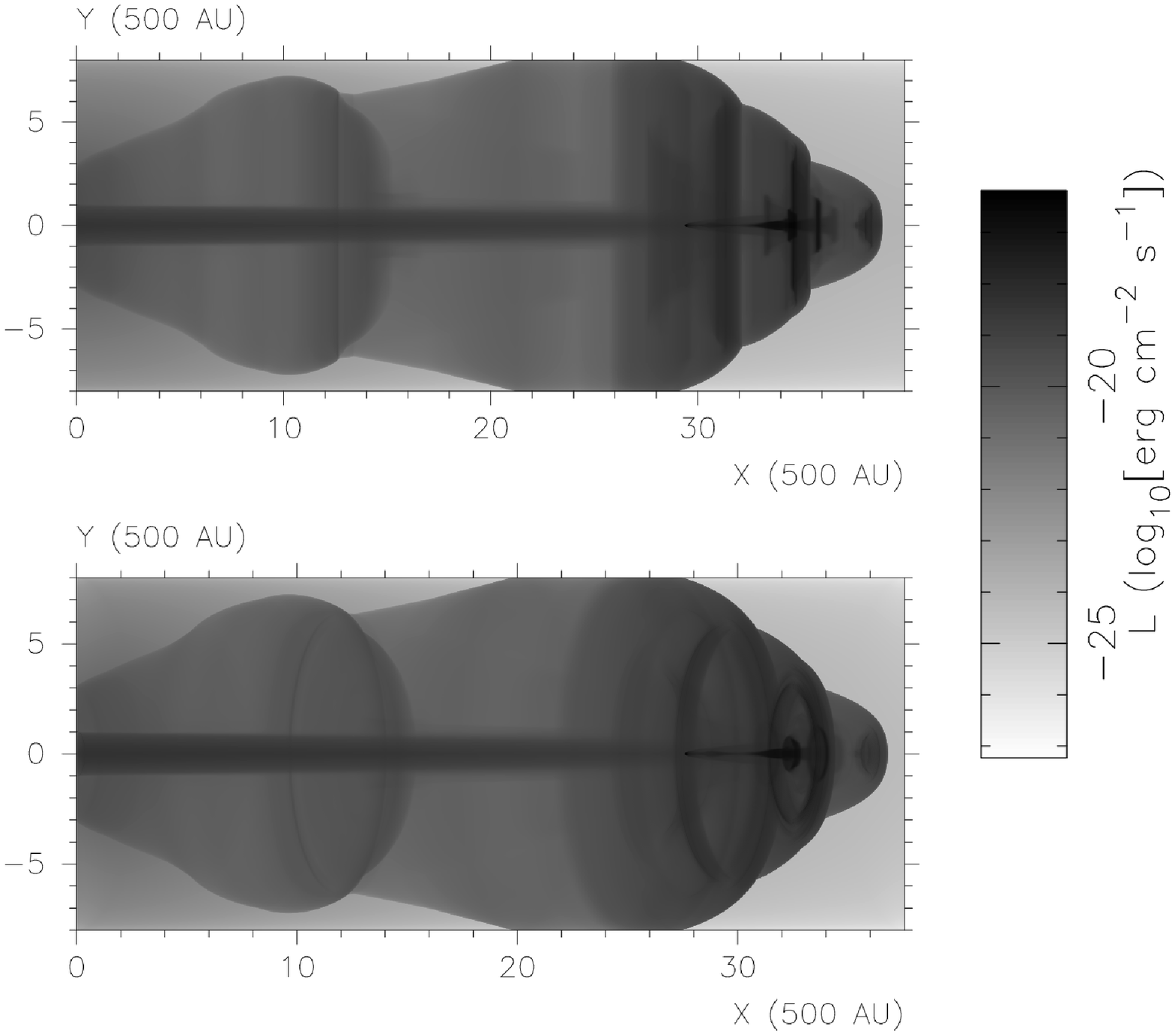}
 \caption{Integrated emission of 2.5D 
          Jet at time $t\simeq 1132 {\rm \ yr}$.
          The data is shown here 
          rotated about the axis
          of symmetry and with
          angles of inclination of the
          symmetry axis with respect
          to the image plain of 
          $\theta=0^\circ$(top), and 
          $\theta=20^\circ$(bottom).
  \label{fig:f5}}
 \end{figure*}

\begin{figure*}[ht]
 \vskip 18pt
 \includegraphics[angle=0,
 		 width=6.0in,
                  keepaspectratio=true,
                  trim= 0 300 0 0
                  clip=true]
    {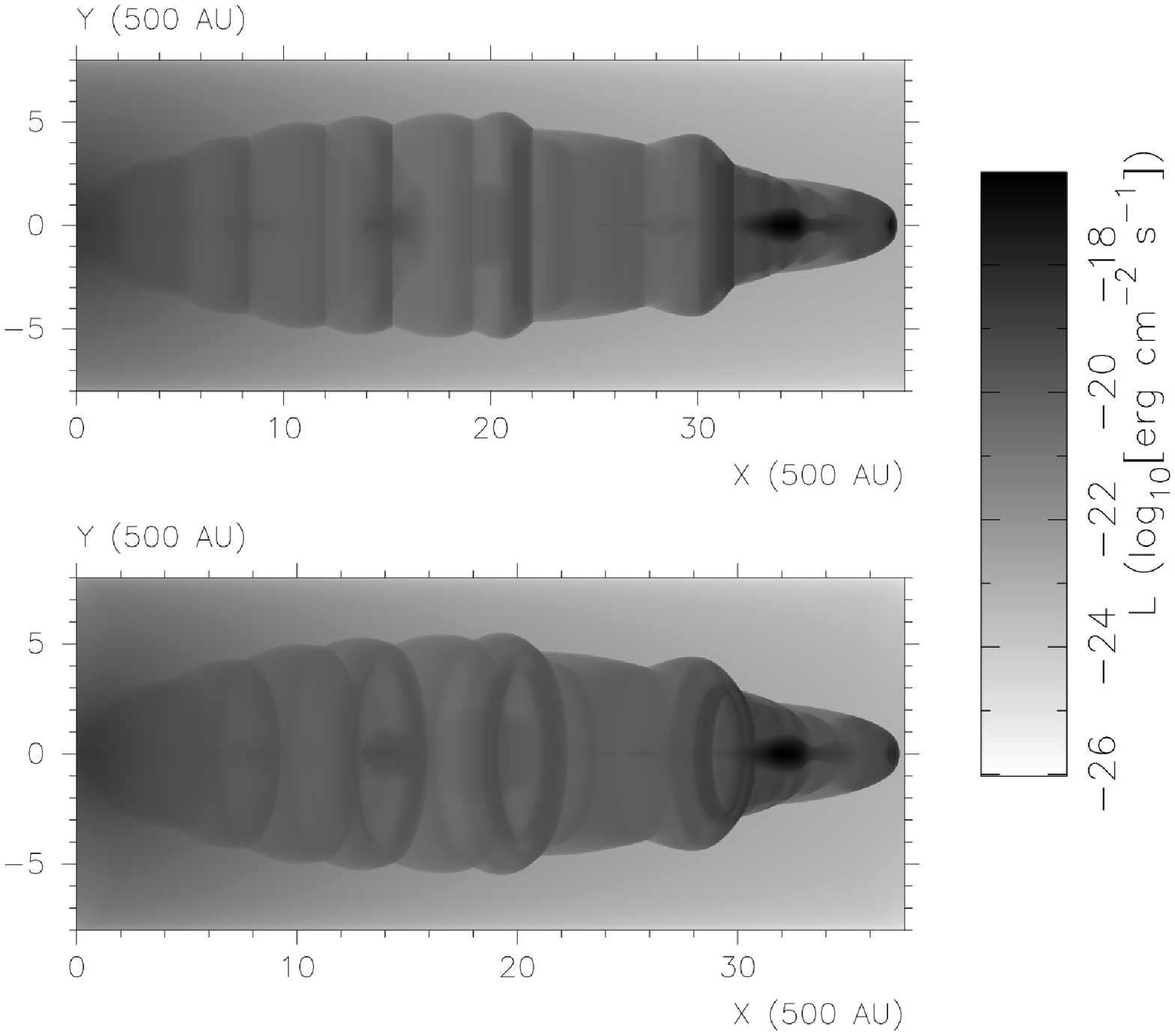}
 \caption{Integrated emission of 2.5D 
          Clump at time $t\simeq 1082 {\rm \ yr}$. 
          The data is shown here 
          rotated about the axis
          of symmetry and with
          angles of inclination of the
          symmetry axis with respect
          to the image plain of 
          $\theta=0^\circ$(top), and 
          $\theta=20^\circ$(bottom).
 \label{fig:f6}}
 \end{figure*}

Results of our simulations are presented in figures \ref{fig:f1}$-$\ref{fig:f6}.
In Figures \ref{fig:f1} and \ref{fig:f2} we present the results of medium 
resolution, (24 cells per radius), 3D simulations of one jet and one clump
respectively.  The length of the domain in these simulations is 20 computational
units with one computational unit corresponding to a physical scale of 500 AU.
(One computational unit is also the value chosen for the radii of the jet and
clump.) In each figure the upper image shows a plot of emission integrated 
along the line of sight which in the case of these figures is perpendicular to the
plane of the image. The lower image in each figure is a plot of the logarithm
of density in a plane coincident with the $x$-$y$ plane.  The images show the 
jet and clump near the end of their respective runs at time 
$t\simeq 498 {\rm\ yr}$ for the case of the clump and at time 
$t\simeq 636 {\rm\ yr}$ for the case of the jet.
The resolution is seen to be sufficient to capture vortex-shedding events in 
both simulations. It is also evident from these images that while one can easily 
distinguish jet from clump in the density maps, the emission maps are quite similar. 
We note that the clump gives rise to a somewhat more collimated flow, while the jet 
bow shock expands laterally at a greater rate than the clump bow shock.  
The jet also lags behind the clump in its forward motion. This is likely due to the 
streamlining that occurs as the head of the clump is reduced in size as material is 
ablated away via its interaction with the ambient medium.
\par
Due to limits on computational resources, it was necessary to impose
limits on the resolution and run time of the 3D simulations from which the 
images in figures \ref{fig:f1} and \ref{fig:f2} are taken. The simulations 
end just as the vortex-shedding events begin to have an interesting effect
on the nebular environment. To explore this stage further we carried out the
second pair of 2.5D simulations mentioned above. In these high-resolution
simulations, the effective resolution was doubled to 48 cells per jet/clump
radius, the length of the domain was doubled, and the transverse dimensions
of the domain were enlarged in an attempt to accomodate the lateral expansion
of the jet/clump bow shock (this latter adjustment was successful only for the
case of the clump).  Results from these simulations are presented in figures
\ref{fig:f3} through \ref{fig:f6}. In figures \ref{fig:f3} and \ref{fig:f4} we
again present images of the logarithm of density for the jet and clump
respectively---this time reflected about the axis of symmetry. Both figures
show the simulations at various stages of the flow. We observe 
that the differences found between the two models in the 3D simulations---i.e.
the faster domain crossing time and the higher degree of collimation exhibited
by the clump---are seen again in these images. The vortex shedding however,
is now captured with greater clarity for both jet and clump, and we begin to see
significant qualitative differences in the manner in which these events 
unfold. In particular we note that shedding events are much more frequent in
the case of the clump. These results mirror those found by Poludnenko, 
Frank \& Mitran (2004).  It is noteworthy that their study used a different 
integration scheme than used here.  AstroBEAR has a number of schemes built
into it and in the Poludnenko study a Wave Propagation scheme was used (LeVeque 1997), 
while here a MUSCL-Hancock method is used. The fact that the basic morphology 
of clumps driving bow shocks dominated by vortex shedding events is recovered 
using both schemes gives us confidence in this aspect of the dynamics.

\subsection{Morphology}
To get a better sense of how the differences between the models might appear
observationally, we present integrated emission maps for the jet and clump 
respectively in figures \ref{fig:f5} and \ref{fig:f6}.  These figures were
produced by calculating the effect on the line-of-sight emission resulting from
a rotation of the cylindrically symmetric data set about the axis of
symmetry. The intensity shown, which does not distinguish among cooling lines,
was determined according to:
\begin{equation}
I_{i,j,k} = \Sigma_k n_{i,j,k}^2\Lambda(T_{i,j,k}),
\end{equation}
where $i$,$j$, and $k$ refer to the $x$,$y$ and $z$ directions in the final data cube created 
by rotating $n(r,z)$ and $T(r,z)$ about the axis of symmetry, and $\Lambda$ is 
the cooling function. The images shown correspond to the final frames in each of 
figures \ref{fig:f3} and \ref{fig:f4} respectively. Each figure provides two views 
of the data:  one in which the angle of inclination of the symmetry axis with respect 
to the image plane, $\theta$, is $0^\circ$; and one in which it is $20^\circ$. 
One difference between the jet and clump cases appears in the shape of the head 
of the bow shock. A clump has a finite reservoir of mass which interacts with the 
ambient medium. As the clump propagates down the grid, it drives a (bow) shock wave into 
the ambient medium. A second shock passes through the clump heating and 
compressing it. When cooling is present this "transmitted shock'' first leaves the 
clump flattened. As material is then ablated away via the interactions with the ambient 
medium the remaining clump material becomes dense and streamlined in the direction of 
propagation.  At later times in the simulation the dense core of the clump drives a 
{\bf V}~-~shaped bow shock head.  In the case of a jet the situation is different.  The 
jet head drives a bow shock into the ambient medium and transmitted shock, called a jet 
shock, propagates back into the jet material.  Decelerated jet material flows transverse 
the these shocks inflating a cocoon behind the wings of the bow shock. Unlike the clump 
however, there is always more high speed material behind the jet shock/bow shock pair 
to resupply the interaction. Thus with material continuously flowing into the cocoon,
the bow shock head remains wider and takes on a flatter more {\bf U}~-~shaped configuration.
Such a distinction between {\bf V}~-~and {\bf U}~-~shaped flows may be important in 
comparing with observations.  We note that both the 2.5D and 3D simulations both show 
this difference.  We note however that the axial symmetry will tend to enhance features 
on the axis.  Our 3D runs do not yet have the resolution to accurately track the 
break-up of the clump.  Thus the {\bf V} and {\bf U} bow shock head distinction must 
be considered less than conclusive and await further study.

Vortex shedding provides another morphological distinction.  In the case of the clump, 
the relatively frequent shedding events have led to a series of thin, irregularly spaced 
rings of enhanced intensity centered about the symmetry axis.  These are reminiscent of 
the ring-like structures observed in some collimated  PPN outflows 
[see for example Trammell \& Goodrich (2002)].  The shedding events occurring in the jet 
simulation lead to similar structures, but these are less frequent and somewhat more 
band-like in character. The qualitative differences in the manner in which these 
rings form in the outflows depending on whether one models them as jets or clumps might 
suggest a means of distinguishing between the two models in observations. One must, 
once again, be careful not to over-interpret these results due to limits on the resolution 
and the fact that these simulations are 2.5D. We thus conclude that the  high-resolution 2.5D 
simulations lend weight to 
the assertion that clumps and/or jets can account for observed ringed structures, but neither 
can be ruled out as a model for the collimated outflows observed in the environments of
PPNs. In the meantime, we note that this conclusion in itself is important
with respect to PPN studies as we will discuss in the last section. 

It is also noteworthy that Lee \& Sahai (2003) attempted to model the rings via a pulsed jet.  
Each ring became associated with an ``internal working surface'' where faster moving material swept 
over slower moving material.  The internal shocks lead to transverse motions of shocked material which impinge
upon the bow shock.  As might be expected, the strength of the emission from 
these shocks decreased as the pulse traveled down the length of the beam. 
Such dimming of the rings with distance from the source is not what is observed 
in CRL 618.  The clump on the other hand produces the opposite kind of pattern,
as ablation events on the clump lead to rings that are bright closer to the head
of the bow shock.  

\subsection{Kinematics}
In addition to comparisons of morphology, it is also important to consider
the flow kinematics, since observed flows are often seen to exhibit 
``Hubble-flow'' characteristics---that is, the flow velocity is observed to
increase linearly with respect to distance from the flow origin. To address this 
issue we present in figure \ref{fig:f7}, plots of the $x$-component of the flow 
velocity averaged over the directions transverse to the flow $\left<v_x\right>$.
Velocity tracers were not used in our simulations. In order, therefore,
to differentiate between mildly perturbed ambient gas, and 
gas that is fully involved in the flow, values of velocity $\lesssim 0.01 v_0$ were ignored,
where $v_0$ is the initial velocity of the clump or jet gas.
The top row of the figure shows, from left to right, the results for the 3D jet
and clump respectively while the bottom figure shows the results for the 2.5D 
jet and clump. The data for these plots are taken from 
times near the end of the simulation when the flows have crossed most of the 
domain. In the two plots involving clumps, there are 
large regions of the flow for which the variation of velocity with 
distance is roughly linear.  The jets on the other hand,  
fail to model this behavior altogether.
Comparison of the 3D clump plots to the 2.5D case yields an interesting result. 
Recall that the first vortex-shedding events in both clump and jet were observed 
to occur in the 3D simulations shortly before the 
end of the simulation. Because of this they do not have time to perturb the flow in 
a way that might be noticeable in these plots. However, when we examine this phenomenon 
kinematically in the extended spatial domain allowed by the 2.5D 
simulation, we find that while the vortex shedding events, appear to perturb 
the kinematics of both jet and clump,
these perturbations do not alter the overall qualitative 
character of the flows in either case.

\begin{figure*}[ht]
 \vskip 18pt
 \includegraphics[angle=0,
 		 width=6.0in,
                  keepaspectratio=true,
                  trim= 0 300 0 0
                  clip=true]
    {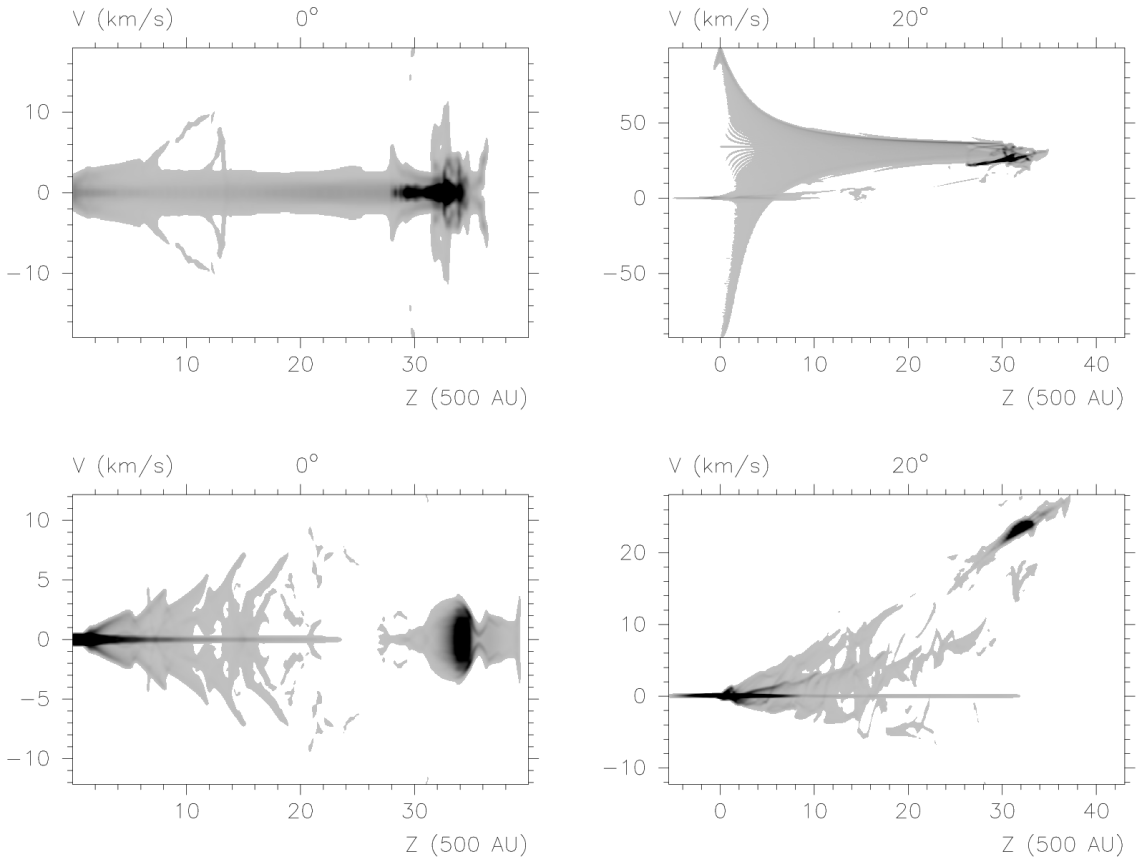}
 \caption{Position-velocity diagrams
          at time $t\simeq 1082 {\rm \ yr}$
          for the 2.5D Jet (top row)
          and the 2.5D Clump 
          (bottom row)
          assuming inclination
          angles of $\theta=0^\circ$(left)
          and $\theta=20^\circ$(right).
 \label{fig:f8}}
 \end{figure*}

To further examine the kinematics of our simulated flows, we have also produced 
a set of synthetic position-velocity (PV) diagrams for the 2.5D simulations. 
These are presented in figure \ref{fig:f8}.  Once again, as in figures \ref{fig:f5}
and \ref{fig:f6} we present our results in pairs corresponding to values 
of $0^\circ$ and $20^\circ$ for the angle of inclination, $\theta$, of the flow symmetry 
axis with respect to the image plane. These plots were produced by calculating
the velocity structure along the line of sight and with the ``slit''
placement taken to be along the projected axis of symmetry.  Results for the jet 
are given in the first row of the figure, and results of the clump are given in
the second row. In these images, the difference between the clump and jet are 
even more striking. For either angle of inclination, the velocity structure of 
the jet cannot be said to be even approximately 
Hubble-like.  The clump however, continues to exhibit line-of-sight velocity 
structure indicative of a linear increase with distance along the projected 
direction of flow. The effect is particularly apparent in the case of the flow 
which is inclined with respect to the image plane. These results, and those of 
figure \ref{fig:f7} suggest that it may be possible to distinguish outflows 
from steady jets from those from explosive events through a 
careful examination of their kinematics.

One caveat which must be considered in these results is the role of emission.
Because our models do not track emission from individual species we cannot 
separate the emission at the bow shock from that within the jet 
or from the shocked jet material.  In Ostriker et al. (2001), a model for the 
emission from a jet-driven bow shock was presented which showed a characteristic 
spur pattern in synthetic PV diagrams.  The spur exhibits a rapid drop in velocity 
away from the tip of the bow shock.  Lee \& Sahai (2003) found a range of patterns  
in their jet simulations which in some cases took on the spur morphology.  Thus 
our results are suggestive of the differences between jets and clumps and indicate 
that clumps appear to be better, in general, at recovering quasi-linear increases in velocity 
along the nebular outflow lobe.

\subsection{Kinematic Models}
\label{sec:KinMod}
\begin{figure*}[ht]
 \vskip 18pt
 \includegraphics[angle=0,
  		 width=6.0in,
                  keepaspectratio=true,
                  trim= 0 300 0 0
                  clip=true]
    {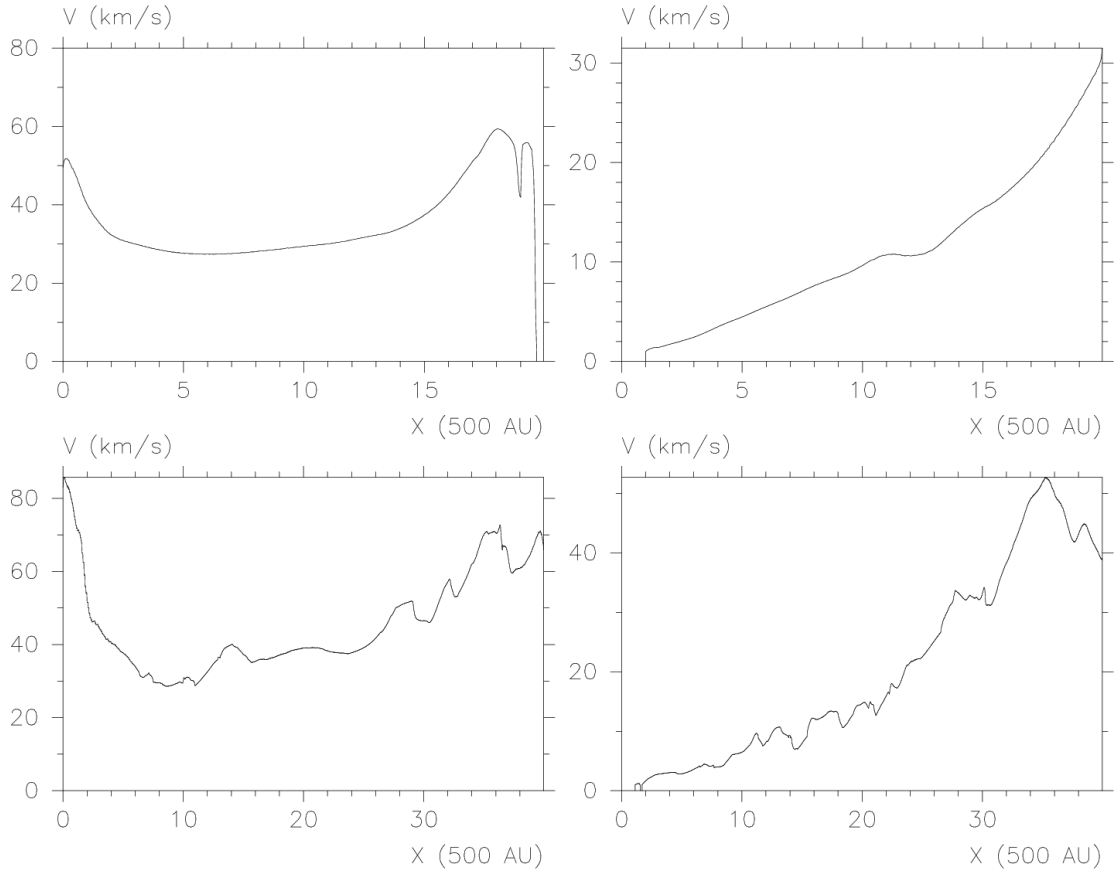}
 \caption{Comparisons of weighted average
          parallel flow velocity,
          $\left<v_x\right>$,
          vs distance from the flow
          origin for clump and jet
          for the 3D 
          simulations(top row)
          calculated at time
          $t\simeq 636 {\rm \ yr}$,
          and the 2.5D 
          simulations(bottom row)
          calculated at time
          $t\simeq 1165 {\rm \ yr}$.
          (See the text for an explanation of the weighting.)
 \label{fig:f7}}
 \end{figure*}

\begin{figure}[ht]
 \includegraphics[width=0.75\textwidth]
  {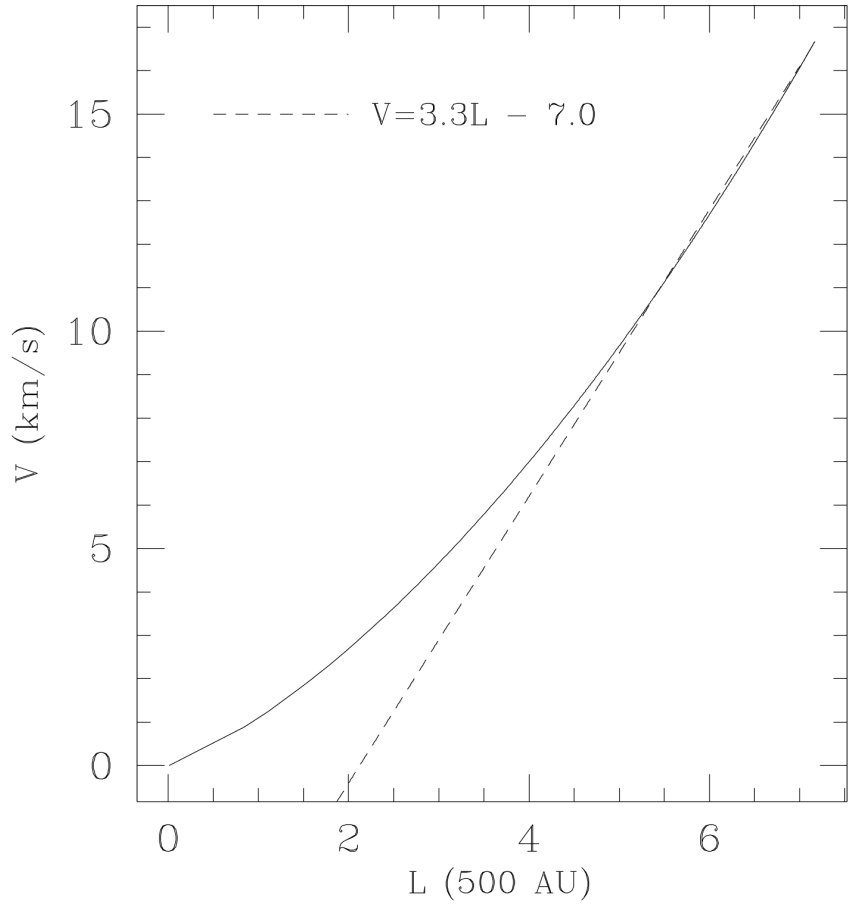}
 \caption{$V(r)$ vs $L(r)$ for $\tilde t=5$. 
See section \ref{sec:KinMod} for an explanation of this figure.
 \label{fig:f9}}
 \end{figure}

In order to interpret our results we consider the time-dependent distortion of
the clump gas during the evolution of the outflow.  Strongly radiative,
hypersonic clouds of any geometry will be rapidly compressed into a thin ballistic 
sheet after ejection by the outflow progenitor.  We therefore consider the motion 
of a cylindrically symmetric disk with surface density $\chi(r)$ and velocity 
$v(r,t)$ where $v(r,0)=v_\circ$ to model the time-dependent evolution of the 
clump gas.  The equation of motion for a differential ring of the disk under
the ram pressure of the ambient gas of density $\rho$ is given by:
\begin{equation}
\rho v^2(r,t) = -\chi \frac{dv}{dt}.
\label{eq:newton}
\end{equation}
Because the outflow bow shock is convex, most of the outflow-entrained
ambient gas will be swept outside of the path of the clump into the
bow shock.  We therefore neglect accretion of ambient material
onto the clump and the kinematics of ambient material ejection in this
model. Because the disk is hypersonic in a strongly cooling
environment, we consider the model disk to be ballistic, neglecting
pressure forces.  For simplicity we also take the density of the ambient
gas to be constant. Thus the equation of motion for a differential ring of clump
gas with radius $r$ integrates to:
\begin{equation}
v=\frac{v_0}{1+\rho v_0 t/\chi}.
\label{eq:eqvkin}
\end{equation}
The distance traversed by the ring is given as:
\begin{equation}
L(\chi,t)=\int_0^t v(t) dt=\frac{\chi}{\rho}\ln\left[1+\frac{\rho v_0 t}{\chi}\right].
\label{eq:eqLkin}
\end{equation}
The quantities $v$, $t$, and $L$, all refer to the same ring. 
What differentiates one ring of material from another is the parameter $r$. 
Now at some late time $t$, we imagine the rings to have been distributed
over the length of the outflow with this distribution depending on $r$. 
For this fixed value of $t$, we are interested plotting the 
velocity of each ring against its corresponding distance. 
The $r$-dependence of $v$ and $L$ enter into the expressions for these
quantities through the surface density $\chi$. We therefore model $\chi$
by assuming that the clump of gas from which our disk formed was initially 
spherical, of constant volume mass density $\sigma$, and compressed 
in such a way that all material within the volume of the clump and lying
along a given line passing through the clump in the direction of its motion 
remains on this line after compression. Then,
\begin{equation}
\chi(r)=2\sigma r_0\sqrt{1-(r/r_0)^2},
\end{equation}
where $r_0$ is the radius the clump/disk.
Introducing the dimensionless quantities:
\begin{equation}
\tilde r= \frac{r}{r_0},
\end{equation}
\begin{equation}
\tilde v = \frac{v}{v_0}
\end{equation}
\begin{equation}
\tilde t = \frac{\rho v_0 t}{2\sigma r_0}
\end{equation}
and,
\begin{equation}
\tilde L = \frac{\rho}{2\sigma r_0}L,
\end{equation}
our parametric equations are:
\begin{equation}
\tilde v=\left[
1+\frac{\tilde t}{\sqrt{1-{\tilde r}^2}}
\right]^{-1},
\end{equation}
and
\begin{equation}
\tilde L = \sqrt{1-{\tilde r}^2}\ln
\left[
1+\frac{\tilde t}{\sqrt{1-{\tilde r}^2}}
\right].
\end{equation}
The value of $\tilde t$ is chosen by assuming $\sigma\gtrsim 2\rho$, and by noting that at late times
$v_0 t\gtrsim r_0$. In our 2.5D simulations we have $v_0t/r_0 \sim 4\tilde t \simeq 20$ making 
$\tilde t\simeq 5$. Figure \ref{fig:f9} shows a plot of $v$ vs $L$ in astronomically 
relevant units with this choice of $\tilde t$. For purposes of comparison,
a line with an appropriately chosen slope and intercept is plotted as well.
In spite of the simplicity of our analytical model, a comparison of this plot with the 
upper-right-hand plot of figure \ref{fig:f7} 
reveals good agreement between the two. Both curves exhibit a small
concave curvature for small $L$ while becoming increasingly 
linear with increasing L (the downward turn in the 2.5D simulation-based
plot results from extending the plot into regions not yet reached by the clump).
We also find that the range of $L$-values over which the curve shows linear
behavior increases for increasing $\tilde t$, implying
that a correlation between the kinematical ages
of PPN outflows and the extent to which they are observed to exhibit Hubble-flow.  We take this
calculation as further evidence that our simulations are accurately capturing the
dynamics of the flow and the relevance of bullet models for PPN.

\clearpage


%
%
%
%
 

 
 %
 %
 
%
\clearpage
\section{Discussion and Conclusions}
\label{sec:Discussion}
In this paper we have examined the results of two pairs of simulations intended
to model the gross morphological and kinematical properties of PPNs. Our primary
purpose was to ask not if we could distinguish between jet and clump models, 
but instead to ascertain if clump models could perform equally well at recovering 
these properties. Below we explain the justification for this more explicitly and 
present the conclusion of our study. As was discussed in the introduction, MHD models of 
PN shaping have been explored by a variety of authors and in a variety of forms.  
As we will explain below, it is the imperative of MHD models, particularly 
magneto-centrifugal launch and collimation scenarios, which motivate this paper.

Two distinct classes of model for the magnetic shaping of winds in PNs and PPNs 
have been suggested to date. First there is the Magnetized Wind Bubble (MWB) model, 
originally proposed by Chavalier \& Luo (1994) and studied numerically by 
R\`o\.zyczka \& Franco (1996) and Garcia-Segura et al. (1999).
In these models, an initially weak toroidal magnetic field is embedded in 
a radiatively-driven wind. This configuration has been shown capable of 
accounting for a wide variety of outflow morphologies including highly~-~collimated jets. 
These models clearly demonstrate the importance of magnetic fields. 
They cannot however account for the excess momentum in the flows (Bujarrabal et al. 2001) 
because of the weak fields which simply ride along in a radiatively-driven wind. 
\par
The other class of model invokes so-called Magnetocentrifugal Launching (MCL; 
Blandford \& Payne 1982; Pelletier \& Pudritz 1992).  This paradigm has, for 
many years, been explored as the mechanism driving jets in young stellar 
objects (YSOs), micro-quasars and active galactic nuclei (AGNs). The MCL 
paradigm assumes the presence of a rotating central gravitating object 
(which may or may not include an accretion disk).  In the case of a disk, 
plasma is threaded by a magnetic field whose 
poloidal component is in co-rotation with the disk. Disk-coronal gas is then 
subject to centrifugal force which accelerates the gas flinging it out 
along field lines.  The magnetized plasma eventually expands to a configuration 
where the toroidal component of the field dominates and hoop stresses collimate 
the flow. Thus the MCL paradigm accounts for both the origin of the wind and the 
means of collimation.
\par
The success of the MCL paradigm in modeling jets associated with YSOs and AGNs 
has led some authors to suggest applying the idea in the context of PNs and PPNs 
(Blackman et al. 2001a, Frank \& Blackman 2004,). Most recently it has been shown 
that the observed total energy and momentum in PPNs 
can be recovered with disk wind models using existing disk formation scenarios 
via binary interaction (Frank \& Blackman 2004 and references within). 
\par
Most theoretical investigations of the MCL paradigm assume a steady-state flow. 
Observations suggest however, that acceleration times for the flows are
as much as an order of magnitude shorter than typical kinematical PPN ages 
(Bujarrabal et al. 2001). This implies that the mechanism responsible for the 
observed flows may operate explosively, i.e. the time over which the 
mechanism acts is short compared to the lifetime of the flow. Moreover, it has 
been suggested by Alcolea et al. (2001) that such a 
scenario would also provide the most straightforward explanation for the ``Hubble
law'' kinematics observed in some PPN outflows (Balick \& Frank 2002; 
Bujarrabal, Alcolea, \& Neri 1998; Olafsson \& Nyman 1999). The MCL paradigm can 
act transiently however, when linked with the rapid evolution of its source, as 
for example in the case of the proposed mechanisms for gamma-ray bursts
(GRBs; Piran 2005) and Supernovae (SNe). This scenario has been investigated by a 
number of authors (Klu\'zniak \& Ruderman 1998; Wheeler et al. 2002; 
Akiyama et al. 2003; Blackman et al. 2006). In these scenarios differential
rotation twists an initially weak poloidal field thereby generating and amplifying 
a toroidal field.  When the toroidal component reaches a critical value it drives 
through the stratified layers of the collapsing core carrying trapped material with it. 
The hoop stresses associated with such a field also serve to collimate the flow.

Recently Matt, Frank, \& Blackman (2006) examined numerically a simplified version 
of this idea which was originally suggested for PPN in Blackman et al. 2001b. In 
these studies the authors began with a gravitating core threaded by an initially 
poloidal field set rotating at 10\% of the escape speed within an envelope of 
ionized gas.  As the simulation progressed, the resulting toroidal field was 
sufficiently strong to drive a complete and rapid expulsion of the gaseous 
envelope. Since the initial conditions assumed in Matt, Frank \& Blackman (2006) are 
applicable to either a young PPN or a collapsing protopulsar, it is 
reasonable to ask whether such transient events are occurring in the early 
stages of the formation of PNs, and if such events can serve as well as 
steady-state jets can in accounting for the complex morphologies observed in such 
systems. We note that these classes of model are sometimes referred to as 
``magnetic towers'' or ``springs'' because it is the gradient of toroidal field 
pressure which drives the outflow. Again we note that the magnetic fields 
needed for our scenario can be delivered by binary interactions as has been 
demonstrated by Nordhaus \& Blackman 2006 and Nordhaus, Blackman \& Frank 2007.

There is growing evidence to suggest that magneto-centrifugal launch models are appropriate 
for PNs and, more importantly, PPNs (Vlemmings et al. 2006).  
Taken together with the evidence that many PPNs have short acceleration time scales 
for which $\tau_{acc} < .1 \tau_{dyn}$ it suggests that some PPNs may be considered 
to have arisen from explosive, or at least impulsive,
depositions of momentum and energy into surrounding, circumstellar environments.
Together we argue that these lines of evidence suggest that many PPNs may be shaped by 
shells which fragment into clumps rather than multiple jets.

There is a subset of PPNs and young PNs with multiple lobes of roughly similar 
size. These include CRL 2688, CRL 618, IRS 19024+004, IRS 09371+1212, M1-37 
and He2-47. It is natural to try and interpret these structures initially as
resulting from the action of jets.  However, consideration of magneto-centrifugal 
launching models shows this to be unlikely.  In all forms of the model, 
gravitational binding energy is tapped via rotational motions of the central source 
about some axis $\mathbf \omega$, and is converted into outflow kinetic energy using the 
magnetic fields as a ``drive belt''.  The existence of a quasi-stable rotational axis is 
a requirement of the models in order to produce a continuous outflow.  Multiple jets of 
equal length are difficult to imagine in such a scenario as the jets would then each require 
their own rotational engine with separate alignments. Even so-called magnetic tower models 
which drive the jet by winding up an initially  weak poloidal field require a net spin 
axis such that $B_\phi ~ 2 \pi n_\phi B_p$ where $n_\phi$ is the number of turns about 
$\mathbf \omega$.  

The production of multiple bow shocks from clumps or bullets driven by a transient MCL process 
is not as difficult to envision.  In Matt, Frank, \& Blackman (2006), it was shown 
that the static envelope or atmosphere of a star could be entirely driven off of a 
rotating magnetized core.  These models relied on the magnetic tower ``spring'' mechanisms, 
and the envelope becomes compressed into a thin shell which rides at the front of the 
expanding magnetic tower.  Such a thin accelerating shell would be subject to a variety of 
instabilities including the Rayleigh-Taylor, Thin-Shell (Vishniac 1993) and Non-linear 
Thin Shell (Vishniac 1994) modes, all of which would be modified by the presence of 
an ordered magnetic field which would impose a long coherence length onto the resulting flow.
The precise detail of such fragmentation in this situation have yet to be calculated and 
stand as an open problem. Given the impulsive acceleration of a dense, radiatively cooling 
shell into a lower density environment, it is likely that the shell would fragment into a 
number of high mach number clumps directed along the poles (and perhaps the equator, 
see Matt, Frank, \& Blackman 2006 and CRL 6888). A potential challenge to this model 
would be the creation of fragmentation modes that can, in some cases produce roughly 
equivalent clumps in terms of propagation direction on either side of the source as is 
seen in some cases.  Given that caveat however, the ability for explosive magnetic 
tower models already explored in the literature to drive unstable shells makes them 
an attractive means of producing high mach number clumps.  As we have shown, 
these clumps, propelled into the surrounding media then drive bow shocks which do 
at least as good a job as, if not better than, jets in recovering gross 
morphological and kinematic observations. Similar conclusions have recently been
reached by Raga et al. (2007).

In summary, the results presented here add weight to an emerging paradigm in 
which transient (explosive) MCL processes act as the driver for PPN evolution 
in some case. The fact that such magnetic launch mechanisms are already favored 
by some theorists to explain supernovae and gamma-ray bursts (Piran 2005), makes 
all the more compelling the notion that lower energy analogues of the processes 
believed to be occurring during the penultimate stages of massive stars' 
evolution, are also occurring in low and intermediate mass stars.

\acknowledgements
The authors thank Orsola De Marco and Pat Huggins for providing insights in a number of ways.

This work was supported by Jet Propulsion Laboratory Spitzer Space Telescope 
theory grant 051080-001, NSF grants AST-0507519, Hubble Space Telscope theory grant 11251 
and the Laboratory for Laser Energetics.

\end{document}